# Dependences of the Transport Scattering Time and Quantum Lifetime on the Two-Dimensional Electron Gas Density in Modulation-Doped Single GaAs Quantum Wells with AlAs/GaAs Short-Period Superlattice Barriers

A. A. Bykov[a,b], I. S. Strygin[a], A. V. Goran[a], D. V. Nomokonov[a], and A. K. Bakarov[a]

[a] *Rzhanov Institute of Semiconductor Physics, Siberian Branch, Russian Academy of Sciences, Novosibirsk, 630090 Russia*
[b] *Novosibirsk State University, Novosibirsk, 630090 Russia*

The dependences of the transport scattering time $\tau_t$, quantum lifetime $\tau_q$, and their ratio $\tau_t/\tau_q$ on the density $n_e$ of the electron gas in modulation-doped single GaAs quantum wells with AlAs/GaAs short-period superlattice barriers are investigated. The experimental dependences are explained in terms of electron scattering by remote ionized donors with an effective two-dimensional concentration $n^*_R$ and background impurities with a three-dimensional concentration $n_B$. An expression for $n^*_R(n_e)$ is obtained including the contribution of X-valley electrons localized in AlAs layers to the suppression of scattering by the random potential of remote donors. It is shown that the experimentally observed abrupt increase in $\tau_t$ and $\tau_q$ with an increase in $n_e$ above a certain critical value $n_{ec}$ is related to a decrease in $n^*_R$. It is established that the drop in $\tau_t/\tau_q$ observed for electron densities $n_e > n_{ec}$ occurs because scattering by the random potential of background impurities in this two-dimensional system with a decrease in $n^*_R$ limits an increase in $\tau_t$ more considerably than an increase in $\tau_q$.

Since the advent of high-mobility GaAs/AlGaAs heterostructures, they have been the subject of extensive studies and ongoing improvement [1–3]. The high mobility μ of the two-dimensional (2D) electron gas in these heterostructures is obtained by the use of selective doping, i.e., in essence, by the spatial separation of the region where charge-carrier transport takes place from that where doping is introduced. In conventional GaAs/AlGaAs heterojunctions, the chargetransport and doping regions are separated by an undoped AlGaAs layer, called spacer, with the thickness $d_s$. The larger the $d_s$ value, the weaker the scattering by the random potential of the remote dopants and, accordingly, the higher the carrier mobility in the GaAs quantum well (QW). However, an increase in $d_s$ leads inevitably to a decrease in the density of the 2D electron gas. In order to attain low-temperature mobilities $\mu > 1000$ m$^2$/(V s) in modulation-doped GaAs/AlGaAs heterojunctions, the optimum parameters are $d_s \sim 100$ nm and $n_e \sim 3 \times 10^{15}$ m$^{-2}$ [2].

An increase in the density $n_e$ in high-mobility GaAs/AlGaAs heterojunctions upon a decrease in $d_s$ is accompanied by a decrease in the mobility μ; therefore, it is impossible to obtain high electron densities and mobilities in these heterostructures simultaneously. It was suggested that a high-mobility 2D system with a "thin" spacer ($d_s < 100$ nm) and a high electron density ($n_e > 3 \times 10^{15}$ m$^{-2}$) can be implemented using AlAs/GaAs short-period superlattices (SPSLs) as barriers confining a single GaAs QW [4].

A heterostructure of this kind is shown schematically in Fig. 1. In this case, in addition to the factor of the spatial separation of the doping and transport regions, electron scattering by the random potential of dopants is further suppressed owing to the screening action of X-valley electrons confined in the AlAs layers [4-7]. This approach makes it possible to obtain a high-mobility 2D electron gas with a high density in a single QW with SPSL barriers, thereby offering new possibilities for experimental studies of the fundamental properties of low-dimensional semiconductor systems [8-11].

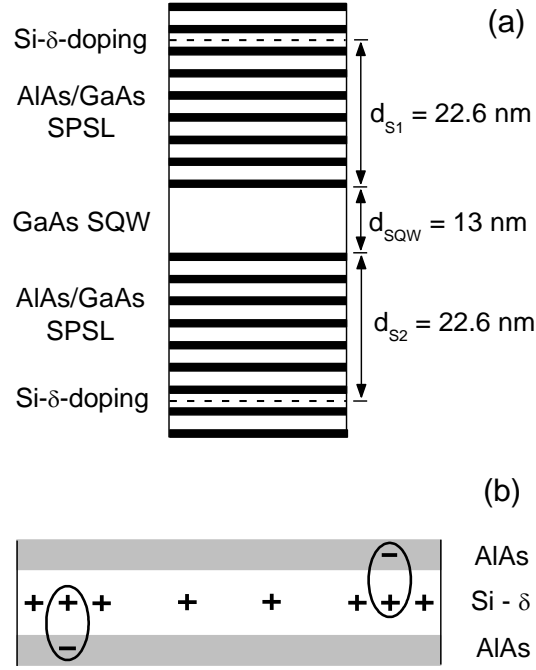

**Fig. 1.** (a) Schematic layout of a single GaAs quantum well confined by AlAs/GaAs short-period superlattice barriers. (b) Enlarged view of a Si-δ-doped layer in a narrow GaAs quantum well with adjacent AlAs layers. Ellipses show compact dipoles formed by positively charged donors in the Si-δ-doped layer and X-valley electrons in AlAs layers [6, 7].

   The investigation of the mechanisms of electron scattering in single GaAs QWs with AlAs/GaAs SPSL barriers is an attractive field of research because the 2D electron gas in these heterostructures can possess a unique combination of parameters such as $n_e$, μ, and the quantum lifetime $τ_q$: the density of electrons $n_e$ is considerably higher and their mobility μ is considerably lower than they are in modern ultrahigh-mobility heterostructures [3, 6, 7]. At the same time, although the mobility is not extremely high, the time $τ_q$ in single QWs with SPSL barriers is quite long, which, in combination with a high density $n_e$, means that fundamental results of general significance can be obtained in these 2D systems [8–11].

   To date, it was shown that impurity scattering in these modulation-doped single GaAs QWs is effectively suppressed by X-valley electrons localized in the SPSL barriers near Si-δ-doped layers [4]. It was also established that no shunt conduction along the doped layers occurs in single QWs with SPSL barriers. A quite unusual relation between the electron mobility and density was found in these heterostructures: the mobility grows abruptly when the density increases above a certain critical value [4]. The dependence μ($n_e$) was qualitatively explained by the screening action of X-valley electrons, but quantitative interpretation of this striking experimental result is still missing.

   The role of X electrons in the suppression of scattering by the random potential of charged impurities in modern modulation-doped GaAs/AlGaAs heterostructures with "thick" spacers and ultrahigh mobilities was recently analyzed analytically and numerically in [6, 7]. Similar to traditional heterojunctions, the spacer in these heterostructures is formed by an undoped AlGaAs layer, while free electrons are supplied by a δ-doped narrow GaAs QW with adjacent AlAs layers [3, 6, 7].

A schematic layout of a section of the Si-δ-doped layer in a narrow GaAs QW with adjacent AlAs layers is shown in Fig. 1b. In the context of the proposed model, X-valley electrons form compact dipoles with ionized donors, which leads to a decrease in the concentration of positively charged impurities and, therefore, to the reduction of the random scattering potential. On the basis of the results of [6, 7], we suggest a method for the analytical description of the dependence of the ionized-donor concentration on $n_e$ using a model function that allows us to quite accurately describe the experimental dependences of $\mu(n_e)$ originally obtained in [4, 5].

Apart from limiting the mobility $\mu = e\tau_t/m^*$, electron scattering in nonideal 2D Fermi systems leads also to the quantum-mechanical broadening $\Gamma = \hbar/2\tau_q$ of single-particle electron states; here, $m^*$ is the electron effective mass and $\tau_t$ and $\tau_q$ are the transport scattering time and quantum lifetime, respectively. Generally, $\tau_t$ and $\tau_q$ are unequal and are defined by the relations

$$1/\tau_q = \int P(\theta)d\theta, \qquad (1)$$
$$1/\tau_t = \int P(\theta)(1-\cos\theta)d\theta, \qquad (2)$$

where $\theta$ is the angle between the Fermi wave vectors before and after electron scattering and $P(\theta)$ is proportional to the probability of scattering at an angle $\theta$. According to Eqs. (1) and (2), $\tau_q$ is determined by the processes of scattering at any angle $\theta$, while $\tau_t$ is determined primarily by large-angle scattering processes owing to the factor $1 - \cos\theta$.

The time $\tau_t$ in high-mobility heterostructures is determined by several scattering mechanisms [12–17]. The low-temperature transport scattering time in a 2D electron gas with high values of $\mu$ and $n_e$ can be written as

$$\tau_t = 1/(1/\tau_{tR} + 1/\tau_{tB}). \qquad (3)$$

Here, $\tau_{tR}$ and $\tau_{tB}$ are the transport times determined by scattering by remote ionized donors and charged background impurities, respectively. These times are determined by the theoretical expressions [13, 15, 16]

$$\tau_{tR} = (8m^*/\pi\hbar)(k_F d_R)^3/n^*_R, \qquad (4)$$
$$\tau_{tB} \cong (2\pi\hbar^3/m^*)(2\varepsilon_0\varepsilon/e^2)^2 k_F^3/n_B, \qquad (5)$$

where $k_F = (2\pi n_e)^{0.5}$, $d_R = (d_S + d_{SQW}/2)$, $d_{SQW}$ is the thickness of the single QW, $n^*_R$ is the effective 2D concentration of remote ionized donors, $\varepsilon_0$ is the permittivity of free space, $\varepsilon$ is the relative permittivity of the single QW and SPSL barrier layers, and $n_B$ is the three-dimensional concentration of charged background impurities.

The time $\tau_q$ in single QWs with SPSL barriers is mainly determined by small-angle scattering [4, 5]. In this case, this time can be written as [13, 16]

$$\tau_q \cong \tau_{qR} = (2m^*/\pi\hbar)(k_F d_R)/n_R^*, \qquad (6)$$

where $\tau_{qR}$ is the quantum lifetime determined by scattering by the random potential of remote impurities. The 2D concentration of remote ionized donors in Eqs. (4) and (6) is their effective concentration $n^*_R$, which includes a change in the concentration of remote ionized donors at their binding with X-valley electrons into compact dipoles. In this way, the above expressions for $\tau_{tR}$ and $\tau_{qR}$ include the screening of the random potential of ionized remote donors by X-valley electrons localized in AlAs layers. We take this approach because no theory describes the dependences $\tau_{tR}(n_e)$ and $\tau_{qR}(n_e)$ in the entire range of variation of the fraction $f$ of positively charged remote donors occupied by X electrons [6, 7].

The heterostructures under study were grown by molecular beam epitaxy on GaAs (100) substrates and contained a single GaAs QW with a width of $d_{SQW}$ = 13 нм. The QW was confined by AlAs/GaAs SPSL barriers [4]. Free electrons were supplied by two Si-δ-doped layers located in two narrow GaAs QWs within the SPSL barriers at a distance of $d_S$ = 22.6 нм from the interfaces of the single QW. The Hall concentration and mobility in the investigated samples at a temperature of $T$ = 4.2 K were $n_H \approx 9.3 \times 10^{15}$ m$^{-2}$ and $\mu \approx 109$ m$^2$/(V s), respectively. Measurements were carried out at $T$ = 4.2 K in magnetic fields $B$ < 2 T on Hall bars with a width of $W$ = 50 μm and a length of $L$ = 250 μm (see the inset of Fig. 2a). To measure the dependences $\rho_{xy}(n_e)$ and $\rho_{xx}(n_e)$, the Hall bars were also furnished with TiAu Schottky gates. The resistivities $\rho_{xy}$ and $\rho_{xx}$ were measured by passing an alternating current $I_{ac}$ with a frequency lower than 1 kHz and an amplitude less than 1 μA.

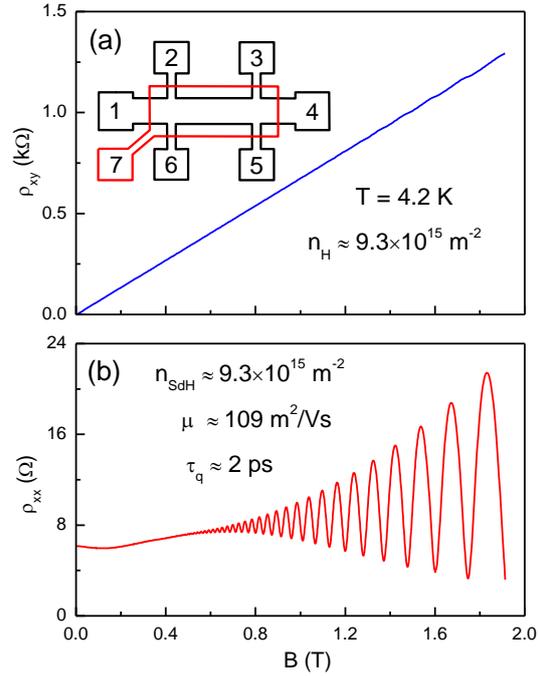

**Fig. 2.** (a) Experimental dependence $\rho_{xy}(B)$ at $T$ = 4.2 K for $V_g$ = 0. The inset shows schematically the Hall bar: (*1–6*) ohmic contacts to the 2D electron gas and (*7*) the Schottky gate (field electrode). (b) Experimental dependence $\rho_{xx}(B)$ at $T$ = 4.2 K for $V_g$ = 0.

Figure 2 shows the experimental dependences $\rho_{xy}(B)$ and $\rho_{xx}(B)$. The dependence $\rho_{xy}(B)$ is linear in the range of magnetic fields $B$ < 1.5 T. The slope of this dependence is determined by the Hall concentration, which makes it possible to determine its value $n_H = B/e\rho_{xy}$. The deviation $\rho_{xy}(B)$ of from the linear dependence in fields $B$ > 1.5 T is caused by Landau quantization, which leads to Shubnikov–de Haas (SdH) oscillations. The period of SdH oscillations in the inverse magnetic field is determined by the density of the 2D electron gas in the GaAs QW: $n_{SdH} = 2(e/h)f_{SdH}$, where $f_{SdH}$ is the frequency of SdH oscillations. The amplitude of SdH oscillations is determined by the relation $\Delta\rho_{xx} = 4\rho_0 X(T)\exp(-\pi/\omega_c\tau_q)$, where $\rho_0 = \rho_{xx}(B = 0)$, $X(T) = (2\pi^2 k_B T/\hbar\omega_c)/\sinh(2\pi^2 k_B T/\hbar\omega_c)$ is the temperature factor, and $\omega_c = eB/m^*$ is the cyclotron frequency [14]. We used this relation to determine $\tau_q$ from the experimental dependences of $\Delta\rho_{xx}(1/B)$.

Figure 3a shows the dependences of $n_H$ and $n_{SdH}$ on the gate voltage $V_g$. The concentration $n_H$ was calculated using the value of $\rho_{xy}$ in a magnetic field of $B = 0.5$ T, while $n_{SdH}$ was obtained from the period of SdH oscillations. In the presence of a conductive shunt, $\rho_{xy}$ will be determined by the total density of free charge carriers in the single GaAs QW and in the δ-doped layers within the SPSL (shunt). In this situation, $n_H$ will be equal to the sum of the electron densities in the single GaAs QW and in the modulation-doped AlAs/GaAs SPSL. This value should be higher than $n_{SdH}$, but we find that $n_H = n_{SdH}$ within an accuracy of ~ 1% in the entire investigated range of $V_g$. Therefore, we can assume that $n_H \approx n_{SdH} \equiv n_e$. This means that the shunt conductivity in the samples under study is much lower than the conductivity of the 2D electron gas in the single GaAs QW and can be ignored [18].

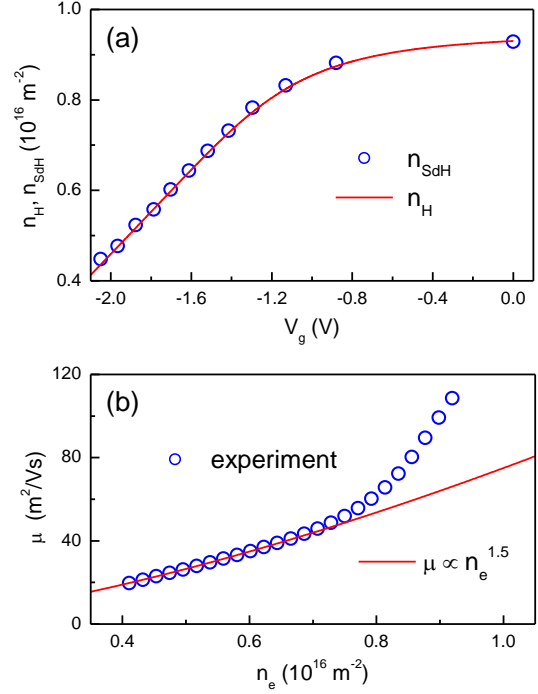

**Fig. 3.** (a) Experimental dependences $n_H(V_g)$ and $n_{SdH}(V_g)$ (solid line and circles, respectively) at $T = 4.2$ K. (b) Mobility μ versus the electron density $n_e$. Circles show experimental data at $T = 4.2$ K; the solid line is calculated according to the relation $\mu \propto n_e^{1.5}$.

Figure 3b shows the behavior of the mobility μ as a function of $n_e$. Circles represent the experimental data, and the solid line corresponds to the theoretical dependence $\mu \propto n_e^{1.5}$. The discrepancy between the experimental and theoretical dependences is qualitatively explained by the screening role of X-valley electrons localized in the AlAs layers [4, 5]. However, quantitative analysis of the experimental dependence of the mobility on the electron density in the region where this dependence deviates from the law $\mu \propto n_e^{1.5}$ is currently unavailable [4–7].

The investigated single GaAs QW with AlAs/GaAs SPSL barriers is symmetrically doped. The concentration of Si donors in each of the δ-doped layers, located at a distance $d_S \equiv d_{S1} = d_{S2}$ from the interfaces of the single QW, was ~ $2 \times 10^{16}$ m$^{-2}$. This means that the total concentration of positively charged Si donors is $n_R \leq 4 \times 10^{16}$ m$^{-2}$. For $V_g = 0$, the concentration is $n_{SdH}$ ~ $9.3 \times 10^{15}$ m$^{-2}$. It follows that the electron density in the AlAs layers adjacent to the δ-doped layers at $V_g = 0$ is on the order of $3 \times 10^{16}$ m$^{-2}$. Since $n_{SdH} \approx n_H$, all these electrons are localized, and their contribution to the conductivity can be disregarded. A model of this localization was recently proposed in [6, 7].

When the voltage $V_g$ is applied to the gate, this leads not only to a change in $n_{SdH}$ but also to a change in the density of X electrons localized in the AlAs layers located near the upper δ-doped layer. At the same time, the situation in the AlAs layers adjacent to the lower layer remains unchanged.

The dependence $n_e(V_g)$ shown in Fig. 3a has two characteristic regions. In the range of $V_g$ from 0 to –1.5 V, this dependence is nonlinear because the variation of the gate voltage within this range changes not only the electron density $n_e$ in the single QW but also the density of X-valley electrons localized in the SPSL barriers. This means that the screening of the random potential of dopants changes in the nonlinear part of the dependence $n_e(V_g)$, and this potential cannot be considered fixed. In this case, the dependence $\mu(n_e)$ is determined not only by the variation in $k_F = (2\pi n_e)^{0.5}$ but also by the change in the concentration of scattering centers, as well as in correlations in their spatial arrangement and in small-angle scattering events. This is one of the reasons why $\mu(n_e)$ is not described by the law $\mu \propto n_e^{1.5}$ in the range $n_e = (0.65\text{-}0.93)\times 10^{16}$ m$^{-2}$.

For gate voltages $V_g$ from –1.5 to –2.1 V, the dependence $n_e(V_g)$ is linear. Therefore, it can be assumed that the capacitance $C_{17}$ between the Schottky gate and the ohmic contact to the conductive layers of the heterostructure (terminals *1* and *7* in the inset of Fig. 2a) in this range is independent of $V_g$, which is fully consistent with the results and conclusions of [4]. We agree with the authors of [4] that this behavior reflects the constancy of the ionized-donor concentration in this range of gate voltages. Comparing the capacitance measured from the slope of the linear segment of the dependence $n_e(V_g)$ with the geometric capacitance shows that the two values differ in the samples under study by approximately 4%, which corresponds to the experimental error.

Thus, we can conclude that the scattering potential remains unchanged in the linear region of the dependence $n_e(V_g)$. In this case, the behavior of $\mu(n_e)$ is determined only by the change in $k_F$ and is well described by the function $\mu \propto n_e^{1.5}$. The experimental dependence in the range $n_e = (0.4\text{-}0.65)\times 10^{16}$ m$^{-2}$ fully agrees with theoretical relations (4) and (5). In this case, we can assume that the time $\tau_t$ is determined by scattering by two random potential types induced by remote dopants and background impurities. However, we cannot separate the contributions from these two random scattering potentials to $\tau_t$ by comparing the experimental data with theory and calculate the values of $n^*_R$ and $n_B$, since both scattering potentials give the same dependence on $n_e$.

Another parameter that characterizes scattering processes in degenerate electron systems is $\tau_q$. In the 2D system under study, it is mainly determined by small-angle scattering [4, 5]. In this case, $\tau_q \approx \tau_{qR}$, and Eq. (6) can be used to analyze the experimental dependence $\tau_q(n_e)$, which is shown in Fig. 4a by circles. It can be described by Eq. (6) with the parameter $n^*_R = 9.28\times 10^{15}$ m$^{-2}$ only in the range of electron densities $n_e < n_{ec} \approx 6.5\times 10^{15}$ m$^{-2}$. In this case, the density $n_{XR}$ of X-valley electrons localized near the upper δ-doped layer is zero. In the range $n_e < n_{ec}$, the effective concentration $n^*_R$ is constant and attains its highest possible value $n^*_R = n^*_{RI} = 9.28 \times 10^{15}$ m$^{-2}$. In the range of $n_e > n_{ec}$, the gate voltage changes both $n_e$ and $n_{XR}$. In this case, the concentration is $n^*_R = n^*_{RI} - n_{XR}$ and cannot be considered fixed; i.e., a different value of $n^*_R$ corresponds to each value of $n_e$.

Using Eq. (6), we calculated the values of $n^*_R$ corresponding to the experimental values of $\tau_q$ and thus obtained the dependence of the ratio $n^*_R/n^*_{RI}$ on $n_e$. This dependence is shown by circles in Fig. 4b. The solid line in Fig. 4b shows the dependence of the ratio $n^*_R/n^*_{RI}$ on $n_e$ calculated by the expression

$$n^*_R/n^*_{RI} = 1/\{\exp[(n_e - a)/b] + 1\} \equiv f_{ab}(n_e), \quad (7)$$

where $a$ and $b$ are fitting parameters. The quantity $f_{ab}$ is, in essence, the fraction of ionized donors that are not occupied by X electrons.

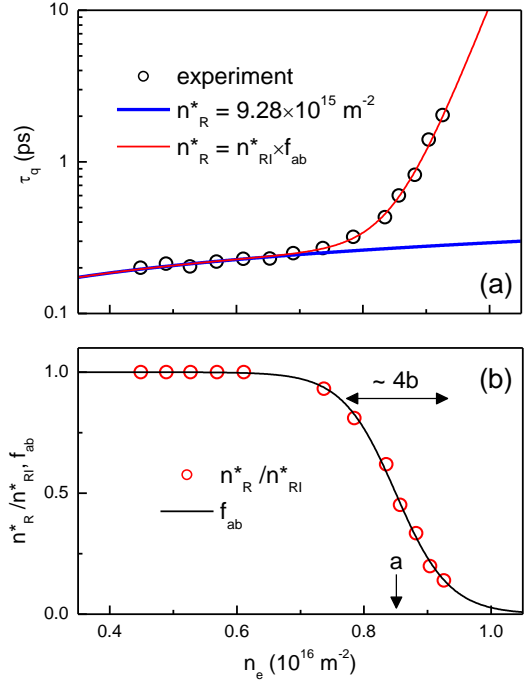

**Fig. 4.** (a) Quantum lifetime $\tau_q$ versus the electron density $n_e$. Circles show experimental data at $T = 4.2$ K; the thick solid line is calculated by Eq. (6) with $n^*_R = 9.28 \times 10^{15}$ m$^{-2}$; and the thin solid line is calculated by Eq. (6) with $n^*_R = n^*_{RI} f_{ab}$, where $n^*_{RI} = 9.28 \times 10^{15}$ m$^{-2}$. (b) Dependences of $n^*_R/n^*_{RI}$ and $f_{ab}$ on $n_e$. Circles show the values of $n^*_R/n^*_{RI}$ calculated from the experimental dependence $\tau_q(n_e)$ using Eq. (6), and the solid line is calculated by Eq. (7) with $a = 8.5 \times 10^{15}$ m$^{-2}$ and $b = 4.2 \times 10^{14}$ m$^{-2}$.

Note that $f_{ab}(n_e = a) = 0.5$, $f_{ab}(n_e = a - 2b) \sim 1$, and $f_{ab}(n_e = a + 2b) \sim 0$; i.e., $f_{ab}(n_e)$ decreases from 1 to 0 within a range of electron densities $\Delta n_e \sim 4b$. The solid line in Fig. 4a shows the dependence $\tau_q(n_e)$ calculated by Eq. (6) where $n^*_R = n^*_{RI} f_{ab}(n_e)$ is substituted for $n^*_R$. Very good agreement between the experimental data and this theoretical dependence is observed. It is noteworthy that $f_{ab} = (1 - f)$, where $f = n_{XR}/n^*_R$ is the fraction of Si donors that have formed compact dipoles with X electrons [6, 7].

Figure 5a shows the experimental and calculated dependences $\tau_t(n_e)$. The calculation was performed under the assumption that $\tau_t = 1/(1/\tau_{tR} + 1/\tau_{tB})$, where $\tau_{tR}$ and $\tau_{tB}$ were calculated by Eq. (4) with $n^*_R = n^*_{RI} f_{ab}(n_e)$ and by Eq. (5) with $n_B = 2.83 \times 10^{21}$ m$^{-3}$, respectively. The calculated curve perfectly agrees with the experimental dependence. It can be seen that, when $f_{ab}(n_e)$ approaches zero with increasing $n_e$, $\tau_t$ is determined mainly by the scattering of 2D electrons by the random potential of background impurities.

Figure 5b shows the experimental and calculated dependences of $\tau_t/\tau_q$ on $n_e$, which are in good agreement. In the region where the dependence is linear, there is complete agreement with the theory of [13]. The ratio $\tau_t/\tau_q$ decreases in the region of $n_e > n_{ec}$ because scattering by the random potential of background impurities in the samples under study limits an increase in $\tau_t$ with increasing more strongly than an increase in $\tau_q$.

We begin our discussion of the presented results with the region of linear dependence $n_e(V_g)$. In this region, scattering by the random potential of remote impurities is taken into account by the effective concentration of positively charged donors $n^*_R = n^*_{RI}$. Since $\tau_t/\tau_q \gg 1$ for $V_g < -1.5$ V, we can disregard the contribution of scattering by background impurities to $\tau_q$ and assume that $\tau_q \cong \tau_{qR}$ is determined mainly by small-angle scattering by remote ionized impurities [14].

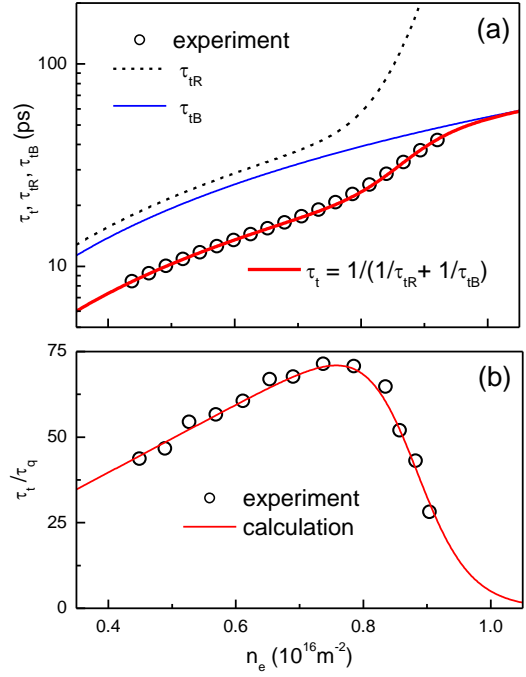

**Fig. 5.** (a) Dependences of $\tau_t$, $\tau_{tR}$ and $\tau_{tB}$ on $n_e$. Circles show experimental data at $T = 4.2$ K; the dashed line is calculated by Eq. (4) with $n^*_R = n^*_{RI} f_{ab}$, where $n^*_{RI} = 9.28 \times 10^{15}$ m$^{-2}$; the thin solid line is calculated by Eq. (5) with $n_B = 2.83 \times 10^{21}$ m$^{-3}$; and the thick solid line is calculated by Eq. (3). (b) Ratio $\tau_t/\tau_q$ versus $n_e$: (circles) experimental data at $T = 4.2$ K and the solid line is the calculation by Eqs. (3) - (7) with $n^*_R = n^*_{RI} f_{ab}$, where $n^*_{RI} = 9.28 \times 10^{15}$ m$^{-2}$ and $n_B = 2.83 \times 10^{21}$ m$^{-3}$.

Comparison of the experimental dependence $\tau_q(n_e)$ in the region of $n_e < n_{ec}$ with that calculated by Eq. (6) gives $n^*_{RI} = 9.28 \times 10^{15}$ m$^{-2}$. The concentration of Si in the upper δ-doped layer is $\sim 2 \times 10^{16}$ m$^{-2}$. Since the concentration in the region of linear dependence of $n_e(V_g)$ is $n_{RX} = 0$, $n^*_{RI}$ should be $\sim 2 \times 10^{16}$ m$^{-2}$, which is inconsistent with the above estimate $n^*_{RI} \sim 10^{16}$ m$^{-2}$. A possible reason for this discrepancy is the correlation of multiple small-angle scattering events, which cannot be treated as independent [14].

The concentration $n_{SdH}$ at $V_g = 0$ is $\sim 10^{16}$ m$^{-2}$, and the total concentration of Si donors in the upper and lower δ-doped layers is $\sim 4 \times 10^{16}$ m$^{-2}$; therefore, $f_{ab} = 0.25$. However, it follows from the experimental dependence $\tau_q(n_e)$ that $f_{ab} = 0.14$ at $V_g = 0$. We also associate this discrepancy with correlations between multiple small-angle scattering events. The determined value $n_B \sim 3 \times 10^{21}$ m$^{-3}$ is noticeably higher than $n_B \leq 10^{20}$ m$^{-3}$ in the samples with $\mu > 1000$ m$^2$/(V s). This indicates a "moderate" quality of the samples under study. However, the presence of background impurities in the single GaAs QW with AlAs/GaAs SPSL barriers hardly reduces the time $\tau_q$ in a 2D electron gas with a high density $n_e$, which is due to the reduction of the random scattering potential of remote donors by X-valley electrons. For this reason, despite the moderate mobility, modulation-doped single QWs with SPSL barriers are widely used to study fundamental quantum phenomena in high-density 2D electron systems, including those with several occupied energy subbands [19–22].

In summary, we have studied the dependences of the transport scattering time $\tau_t$, quantum lifetime $\tau_q$, and their ratio $\tau_t/\tau_q$ on the 2D electron gas density $n_e$ in modulation-doped single GaAs QWs with AlAs/GaAs SPSL barriers. The experimental data have been analyzed under the assumption that the time $\tau_t$ in the 2D system under study is determined by scattering by remote ionized donors with an effective 2D concentration $n^*_R$ and background impurities with the three-dimensional concentration $n_B$, while $\tau_q$ is determined only by scattering by remote ionized donors.

We have proposed an expression for $n^*_R(n_e)$ that includes the contribution of X-valley electrons to the suppression of scattering by the random potential of remote ionized donors. Using this expression, we have separated the contributions to $\tau_t$ from scattering by remote ionized donors and background impurities. We have established the role of background impurities in the processes of 2D electron gas scattering in the investigated modulation-doped GaAs single QWs with AlAs/GaAs SPSL barriers and have estimated the concentration of these impurities. The proposed approach to the analysis of the experimental dependences of $\tau_t$, $\tau_q$, and their ratio $\tau_t/\tau_q$ on the 2D electron gas density $n_e$ have also made it possible to qualitatively explain the drop of $\tau_t/\tau_q$ with increasing observed in [23, 24].

ACKNOWLEDGMENTS

We are grateful to G.M. Min'kov and V.A. Tkachenko for useful discussions.

FUNDING

This work was supported by the Russian Foundation for Basic Research (project nos. 18-02-00603 and 20-02-00309).